\documentclass[runningheads]{llncs}
\usepackage{graphicx}
\usepackage{textcomp}
\usepackage{comment}

\usepackage{color}
\usepackage{graphicx}
\usepackage{amsmath}
\usepackage{amssymb}
\usepackage{booktabs}
\usepackage{verbatim}
\usepackage{import}
\usepackage{multirow}
\usepackage[capitalize]{cleveref}
\usepackage{subcaption}
\usepackage[utf8]{inputenc}
\usepackage[english]{babel}
\usepackage{biblatex}
\usepackage{csquotes}

\usepackage{xspace}
\makeatletter
\DeclareRobustCommand\onedot{\futurelet\@let@token\@onedot}
\def\@onedot{\ifx\@let@token.\else.\null\fi\xspace}

\def\eg{\emph{e.g}\onedot} 
\def\ie{\emph{i.e}\onedot} 
 
 \def\vs{\emph{vs}\onedot}

\makeatother

\newcommand*{\figuretitle}[1]{{\centering \textbf{#1}\par\medskip}}
\usepackage{todonotes}

\begin{document}
\titlerunning{A systematic study of race and sex bias in CNN-based CMR segmentation}

\title{A systematic study of race and sex bias in CNN-based cardiac MR segmentation}

\newcommand*\samethanks[1][\value{footnote}]{\footnotemark[#1]}
\newcommand{\rowstyle}[1]{\gdef\currentrowstyle{#1}%
  #1\ignorespaces
}

\author{Tiarna Lee \inst{1}
\and Esther Puyol-Ant\'on \inst{1} \and 
Bram Ruijsink \inst{1,2}
\and Miaojing Shi \inst{3}\thanks{Joint last authors}
\and Andrew P. King \inst{1}\samethanks{} }
\authorrunning{T Lee et al.}   
\institute{School of Biomedical Engineering \& Imaging Sciences, King\textquotesingle s College London, UK. \and Guy’s and St Thomas’ Hospital, London, UK. \and Department of Informatics, King's College London, UK.}

\maketitle              

\begin{abstract} 
In computer vision there has been significant research interest in assessing potential demographic bias in deep learning models. One of the main causes of such bias is imbalance in the training data. In medical imaging, where the potential impact of bias is arguably much greater, there has been less interest. In medical imaging pipelines, segmentation of structures of interest plays an important role in estimating clinical biomarkers that are subsequently used to inform patient management. Convolutional neural networks (CNNs) are starting to be used to automate this process. We present the first systematic study of the impact of training set imbalance on race and sex bias in CNN-based segmentation. We focus on segmentation of the structures of the heart from short axis cine cardiac magnetic resonance images, and train multiple CNN segmentation models with different levels of race/sex imbalance. We find no significant bias in the sex experiment but significant bias in two separate race experiments, highlighting the need to consider adequate representation of different demographic groups in health datasets.
\keywords{ Segmentation \and Fairness \and CNN \and Cardiac MRI} 
\end{abstract}

\section{Introduction}
\label{sec:intro}

In the field of healthcare, artificial intelligence (AI) is increasingly being used to automate decision-making processes by aiding the diagnosis and analysis of medical images, producing performances that are equal to, or better than, that of clinicians \cite{ShenArtificialReview}. Although the use of AI has improved performances for a broad range of tasks, biases found in the wider world have also been found in these AI models.
For example, commercial gender classification models were found to perform better on lighter-skinned male faces than on darker-skinned female faces~\cite{Buolamwini2018}. This difference was attributed to the lack of representation of women and non-white faces in the publicly available datasets that the models were trained on. 

\begin{figure}[t]
    \centering
    \includegraphics[width=0.55\textwidth]{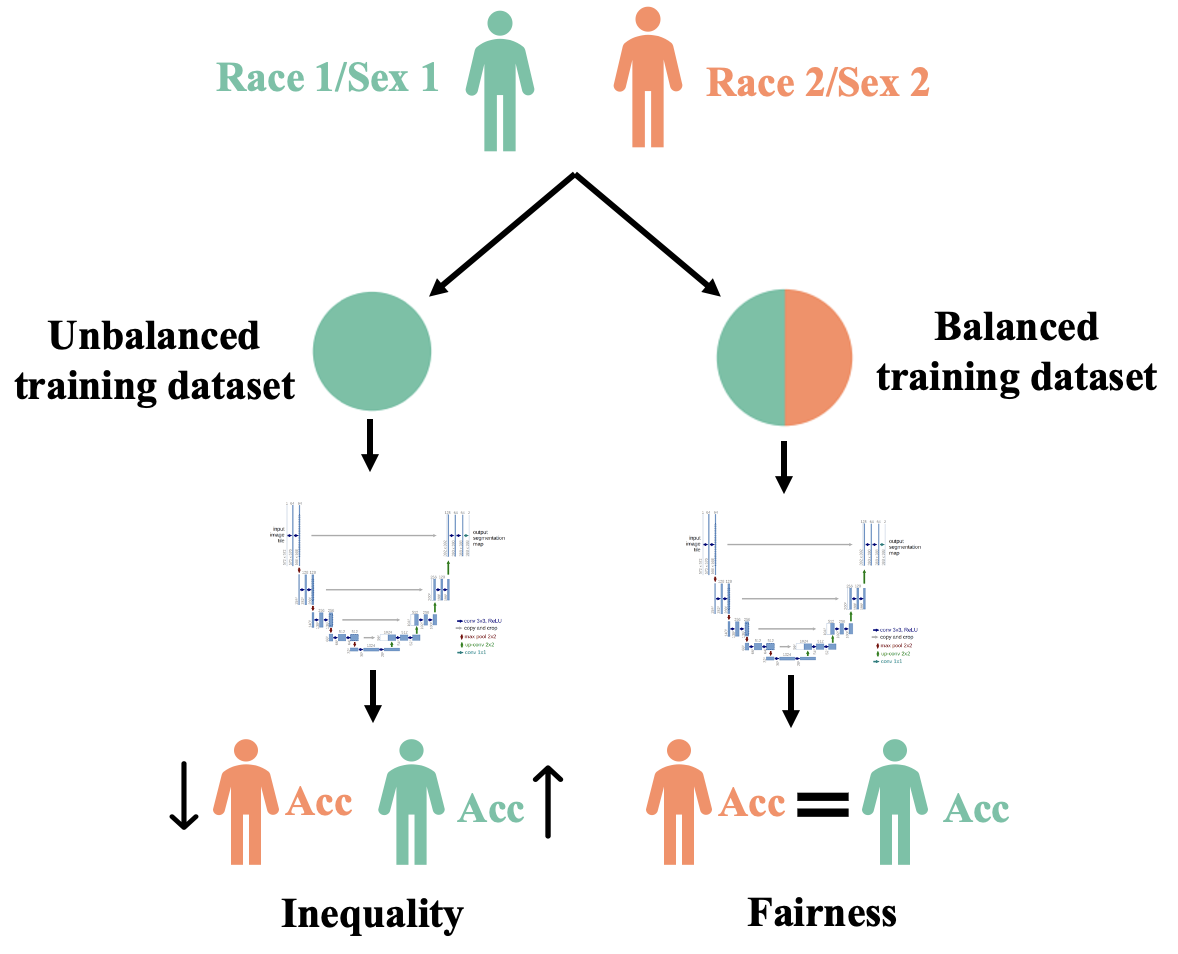}
    \caption{An illustration showing that biases occur in medical image segmentation models as a result of the imbalances of protected attributes in the training datasets.
    }
    \label{fig:dataset_pies}
    \vspace{-3mm}
\end{figure}

In the past few years, research has highlighted the impact of data imbalance on medical imaging tasks. For example, \cite{Abbasi-Sureshjani2020} investigated the impact of demographic imbalance on skin lesion classification models, whilst \cite{Larrazabal2020} performed a systematic study of sex bias for X-ray classification tasks. Furthermore, \cite{10.3389/fcvm.2022.859310, Puyol-Anton2021FairnessSegmentation} demonstrated the existence of racial bias in AI models for segmenting cardiac magnetic resonance (CMR) images. These biases are caused by a combination of training data imbalance and distributional differences between the images acquired from different demographic groups. For example, in X-ray imaging, breast tissue can cause perceptible differences in image characteristics between males and females \cite{Ganz2021}.
Other recent work has suggested that information about protected attribute status can be predicted from a range of medical imaging modalities \cite{Coyner2021,Gichoya2022}, suggesting that these distributional differences are widespread and there is a high potential for bias in AI models applied to medical imaging.



In this work, we perform a systematic study of how training set imbalances in the numbers of subjects from protected groups, such as race and sex, affect the performance of an AI-based segmentation model (see Fig.~\ref{fig:dataset_pies}). We use a dataset of CMR
images to design three experiments: the first studies imbalances in sex by using male and female subjects, the second studies imbalances in race by using white and black subjects, and the third also studies imbalances in race using white and Asian subjects. For each of these experiments, we systematically vary the level of demographic imbalance and measure the performance of the resulting AI models for different protected groups. 
Our key contributions are:
\begin{enumerate}
    \item We significantly extend the preliminary work of \cite{10.3389/fcvm.2022.859310, Puyol-Anton2021FairnessSegmentation} (which only highlighted the presence of race bias) to systematically analyse the impact of sex and race imbalance on CMR segmentation model bias.
    \item We assess the bias in terms of conventional segmentation performance metrics as well as derived clinical measures of cardiac function.
    \item We perform an intersectional study of AI-based segmentation bias, analysing bias effects on groups such as black females, white males, etc.
\end{enumerate}

\section{Materials}
\label{sec:materials}
In this study we use CMR images from the UK Biobank \cite{Peterson2016}.
The images were acquired at four centres across the UK and all centres used the same CMR acquisition protocol. The dataset consists of end diastolic (ED) and end systolic (ES) cine short-axis images from 1,761 subjects. These were a random subset of subjects with no cardiovascular disease or cardiovascular risk factors. We limited our dataset to these subjects to minimise the impact of other potential sources of variation on our analysis (i.e. apart from sex and race).
The demographic data for these subjects were also gathered from the UK Biobank database and can be seen in \cref{tab:subject demographics} for the subjects used in this study. It is important to note that although the UK Biobank uses the term `gender' in their records, in practice the term is more similar to biological `sex' \cite{Tannenbaum2019SexEngineering}.

For each subject, ground truth segmentations for the left ventricular blood pool (LVBP), left ventricular myocardium
(LVM) and right ventricular blood pool (RVBP) were obtained for the ED and ES images via manual segmentation of the LV endocardial and epicardial borders and the RV endocardial border using cvi42 (version  5.1.1,  Circle  Cardiovascular Imaging Inc., Calgary, Alberta, Canada). Each ground truth image was annotated by one expert from a panel of ten who were briefed using the same guidelines. Each expert received a random sample of images from subjects with different sexes and races. The experts were not provided with demographic information about the subjects such as their race or sex.

\begin{table*}
  \caption{Clinical characteristics of subjects used in the experiments. Average values are presented for each characteristic with standard deviations given in brackets. Statistically significant differences, determined using a two-tailed Student's t-test, are indicated with an asterisk * (\(p<0.05\)).}
  \centering
  \resizebox{\textwidth}{!}{
  \begin{tabular}{c|c|cc|ccc}
    \toprule
    Health measure & Overall &  Male & Female & White & Black & Asian \\
     \midrule
     \#Subjects & 1761 & 889 & 872 & 1220 & 238 & 303 \\
    Age (years) & 63.3 (7.9) & 63.7 (8.1) & 62.9 (7.7) & 64.8 (7.6) * & 58.9 (7.0) * & 60.6 (8.0) *\\
    Weight (kg) & 76.5 (14.7) & 83.0 (12.9) *& 69.9 (13.3) * & 76.8 (14.5) & 81.5 (15.9) * & 71.3 (12.9) *\\
    Standing height (cm) & 169.0 (9.2) & 175.3 (6.5) * & 162.5 (6.8) * & 169.8 (9.2) *& 168.7 (9.3) & 165.9 (8.8) * \\
    Body Mass Index (kg) & 26.7 (4.3) & 27.0 (3.6) & 26.5 (4.8) &  26.6 (4.2) &  28.6 (5.0) * & 25.8 (3.6) * \\
    \bottomrule
  \end{tabular}}

  \label{tab:subject demographics}
\end{table*}

\subsection{Experimental setup}
\label{sec:method experimental setup}
To investigate how imbalances in the training datasets affect segmentation performance
for different protected groups, we designed three experiments using datasets with varying levels of imbalance in both sex and race. For each of the three experiments, we created five training
datasets in which the proportions of the subjects varied according to the protected attribute being investigated. For example, when investigating the effect of varying proportions of males and females in the dataset (Experiment 1), the proportion of males and females in Dataset 1 is 0\%/100\%, the proportion in Dataset 2 is 25\%/75\% and so on. We controlled for race in this sex experiment, \ie for each of the five datasets 50\% of the subjects were black and 50\% were white. To investigate the effect of racial imbalance (Experiment 2), the same method as above was applied but here, the five datasets had varying proportions of black and white subjects. Each of the datasets was 50\% male and 50\% female. For both of these experiments, the same test data were used which contained 50\% black and 50\% white subjects, and 50\% male and 50\% female subjects. Experiment 3 also investigated racial bias and was performed using the same method as Experiment 2 but used white and Asian subjects, whilst controlling for sex. The test set for this experiment was comprised of 50\% Asian and 50\% white subjects, and 50\% male and 50\% female subjects.
For each of the experiments, there were 176 subjects in the training set and 84 in the test set.

\section{Methods}
\label{sec:methods}
To assess segmentation performance, the nnU-Net segmentation network ~\cite{Isensee2020} was used to segment the LVBP, LVM and RVBP from the subjects' ED and ES images. The loss function $\mathcal{L}$ was a combination of the cross entropy loss $\mathcal{L}_{ce}$ and the Dice loss $\mathcal{L}_{dice}$,

\begin{equation}
  \mathcal{L} =  \mathcal{L}_{ce} + \mathcal{L}_{dice}.
  \label{eq:loss function}
\end{equation}

\noindent Specifically, $\mathcal{L}_{dice}$ is implemented as 

\begin{equation}
  \mathcal{L}_{dice} = - \frac{2}{|K|} \sum_{k \in K} \frac{\sum_{i \in I} u_i^k v_i^k}{\sum_{i \in I} u_i^k + \sum_{i \in I} v_i^k},
  \label{eq:Dice loss}
\end{equation}

\noindent where $K$ is the number of classes, $I$ is the number of pixels in the training image, $u_i$ is a vector of softmaxed predicted class probabilities at the $i$-th pixel, and $v_i$ is the corresponding one hot encoding of the ground truth class at this pixel.

All models were trained on an NVIDIA RTX A6000. The models were optimised using stochastic gradient descent with a `poly' learning rate schedule, where the initial learning rate was 0.01 and the Nesterov momentum was 0.99. A batch size of 16 was used and the models were trained for 500 epochs. 
During training, data augmentation was applied to the images including mirroring, elastic deformation, gamma augmentation,  rotation and scaling. The nnU-Net was trained using five-fold cross validation on the training set and the resulting five models were used as an ensemble when applied to the test set. Connected component analysis was applied to the  predicted segmentations, with only the largest component retained for each class. The softmax probabilities of the five models were averaged when applying the ensemble to the test set to produce the final predicted segmentations.


\subsection{Model evaluation}
\label{sec:method model evaluation}
Model performance was assessed using the \emph{Dice similarity coefficient} (DSC) which measures the spatial overlap between two sets. For a ground truth segmentation A and its corresponding prediction B, the DSC is given by ${DSC} = \frac{2|A\cap B|}{|A| +|B|}$.

Clinical measures of cardiac function were also calculated for each of the experiments. The end-diastolic volume (EDV) and end-systolic volume (ESV) were first calculated and used to find the ejection fraction (EF) given by $EF= \frac{EDV-ESV}{EDV}$. 


\section{{Results}}
\label{sec:results} 
The overall median DSC scores for the protected group test sets in each of the
three experiments are provided in \cref{tab:average DSC}.  
We also show a visual representation of the three experimental results in \cref{fig:results_boxplots} and performance broken down by region in
\cref{fig:regions_boxplots}. Intersectional analysis of the DSC for each of the experiments can be found in \cref{tab:intersectional_results} and analysis of clinical measures can be found in \cref{fig:clinical_measures}.


\begin{table}[t]
  \centering
    \caption{Overall median DSC for each of the three experiments broken down by the protected attributes used in the experiments. In each experiment we report the results for the respective protected groups and the whole test set. The train percentage signifies the percentages of protected groups used in training in the three experiments. The first and second percentage values correspond with the order of the protected groups in the three experiments, \ie in Experiment 1, 0\%/100\% corresponds with 0\% female, 100\% male; in Experiment 2, 0\%/100\% corresponds with 0\% black, 100\% white, etc. Best results shown in bold.}
  \begin{tabular}{c|ccc|ccc|ccc}
    \toprule
    Train & \multicolumn{3}{c|}{Experiment 1} &  \multicolumn{3}{c|}{Experiment 2}  & \multicolumn{3}{c}{Experiment 3}  \\
    Percentage & Female & Male & All & Black & White & All & Asian & White & All\\
     \midrule
    0\%/100\% & 0.955 & 0.954 & 0.955 & 0.924 & 0.953 & 0.940 & 0.912 & 0.944 & 0.930\\
    25\%/75\% & 0.957 & \textbf{0.963} & 0.959 & 0.956 & \textbf{0.954} & 0.950 & 0.955 & 0.944 & 0.949\\
    50\%/50\% & 0.957 & 0.959 & 0.958 & 0.964 & 0.952 & 0.955 & 0.972 & \textbf{0.944} & 0.954\\
    75\%/25\% & 0.958 & 0.962 & \textbf{0.961} & 0.969 & 0.951 & \textbf{0.960} & 0.974 & 0.936 & \textbf{0.957}\\
    100\%/0\% & \textbf{0.959} & 0.960 & 0.960 & \textbf{0.970} & 0.928 & 0.953 & \textbf{0.974} & 0.913 & 0.943\\
    \bottomrule
  \end{tabular}

  \label{tab:average DSC}
\end{table}

\subsection{Experiment 1: Male \vs Female}
\begin{figure}
\centering
\begin{subfigure}{.3\textwidth}
    \centering
    \figuretitle{Experiment 1}
    \includegraphics[scale=0.3]{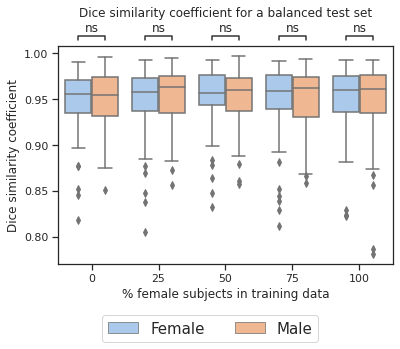}
    \caption{}
    \label{fig:SexDSC}
\end{subfigure}
    \hfill
\begin{subfigure}{.3\textwidth}
    \centering
    \figuretitle{Experiment 2}
    \includegraphics[scale=0.3]{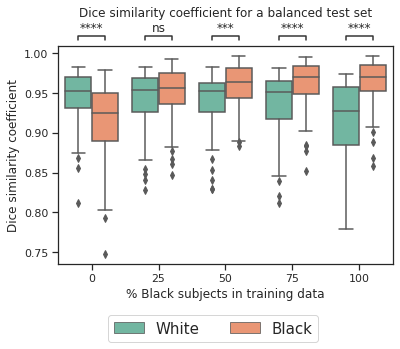}
    \caption{}
    \label{fig:BlackDSC}
\end{subfigure}
    \hfill
\begin{subfigure}{.3\textwidth}
    \centering
    \figuretitle{Experiment 3}
    \includegraphics[scale=0.3]{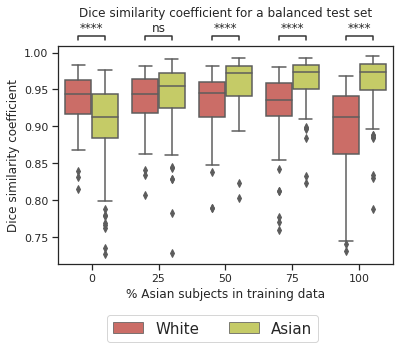}
    \caption{}
    \label{fig:AsianDSC}
\end{subfigure}

\caption{Overall median DSC for the three experiments. Statistical significance was found using a Mann-Whitney U test and is denoted by **** \((p \leq 0.0001)\), *** \((0.001 < p \leq 0.0001)\), ** \((0.01 < p \leq 0.001)\), * \((0.01 < p \leq 0.05)\), ns \(( 0.05 \leq p)\).}
\label{fig:results_boxplots}
\vspace{-3mm}
\end{figure}

The results from Experiment 1 investigating the effect of the imbalance of sex can be seen in \cref{fig:SexDSC}. The performance was reasonably consistent across the five different levels of imbalance. There were no significant differences in overall median DSC for the male and female test sets. However, the intersectional analysis in \cref{tab:intersectional_results} shows that there were significant differences in performance between black and white females, and black and white males, with the black participants achieving significantly higher DSC scores. 
The clinical measures reported in \cref{fig:clinical_measures} also show no statistically significant differences, with the exception that RVESV shows a consistent under-estimation for female subjects.

\subsection{Experiment 2: White \vs Black}
\cref{tab:average DSC} shows that the highest overall median DSC was achieved when the dataset was comprised of 75\% black subjects and 25\% white subjects. The results from \cref{tab:average DSC} and \cref{fig:BlackDSC} show that as the proportion of the protected group increases (\eg 0\% black subjects to 100\% black subjects), the DSC for this group also increases.

Accuracy parity (i.e. approximately equal accuracy between groups) was achieved when the proportion of white subjects was 75\% and the proportion of black subjects was 25\%. By increasing the proportion of black subjects from 0\% to 25\% to achieve accuracy parity, the median DSC for the minority race increased significantly (from 0.924 to 0.956, p$<$0.0001) and increased slightly from 0.953 to 0.954 for the white subjects. This  resulted in a median DSC which was significantly higher overall (0.950 compared to 0.940, p$<$0.0001). Further increasing the proportion of black subjects to 50\% increased the DSC for the black subjects to 0.964.

Interestingly, the results for the clinical measures (\cref{fig:clinical_measures}) show no clear bias. The errors observed, shown by the lengths of the whiskers in the box plots, are consistent across training sets. However, there does appear to be an overestimation of LVESV for both white and black subjects when black subjects are underrepresented in the training set, and an underestimation when white subjects are underrepresented.

\begin{figure*}
\centering
\begin{subfigure}{.3\linewidth}
    \centering
    \figuretitle{Experiment 1}
    \includegraphics[scale=0.3]{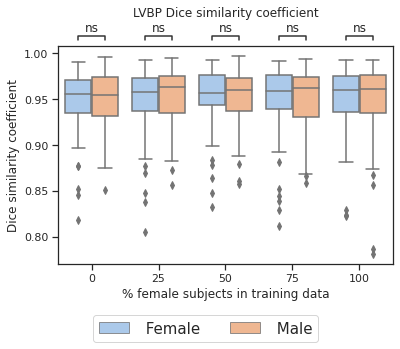}
    \caption{}\label{fig:SexLVBP}
\end{subfigure}
    \hfill
\begin{subfigure}{.3\linewidth}
    \centering
    \figuretitle{Experiment 2}
    \includegraphics[scale=0.3]{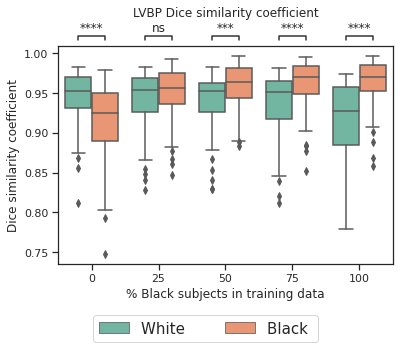}
    \caption{}\label{fig:BlackLVBP}
\end{subfigure}
    \hfill
\begin{subfigure}{.3\linewidth}
    \centering
    \figuretitle{Experiment 3}
    \includegraphics[scale=0.3]{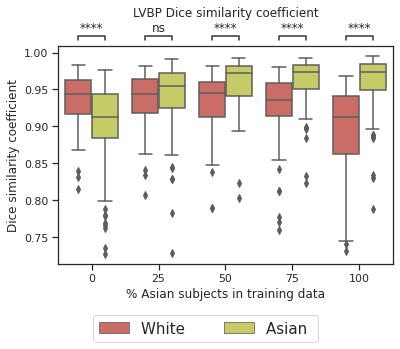}
    \caption{}\label{fig:AsianLVBP}
\end{subfigure}

\bigskip
\begin{subfigure}{.3\linewidth}
    \centering
    \includegraphics[scale=0.3]{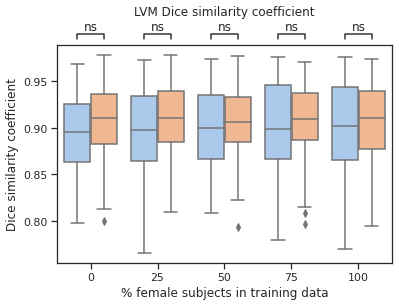}
    \caption{}\label{fig:SexLVM}
\end{subfigure}
    \hfill
\begin{subfigure}{.3\linewidth}
    \centering
    \includegraphics[scale=0.3]{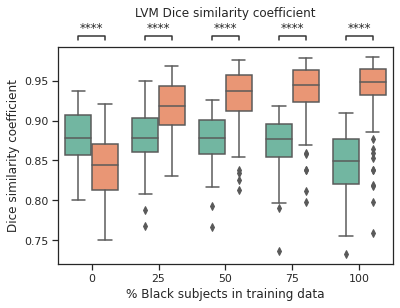}
    \caption{}\label{fig:BlackLVM}
\end{subfigure} 
    \hfill
\begin{subfigure}{.3\linewidth}
    \centering
    \includegraphics[scale=0.3]{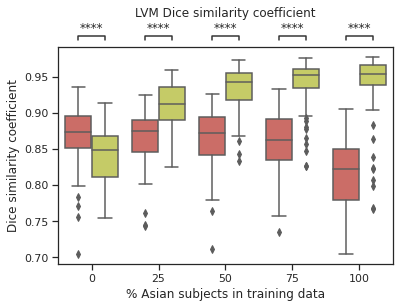}
    \caption{}\label{fig:AsianLVM}
\end{subfigure}

\bigskip
\begin{subfigure}{.3\linewidth}
    \centering
    \includegraphics[scale=0.3]{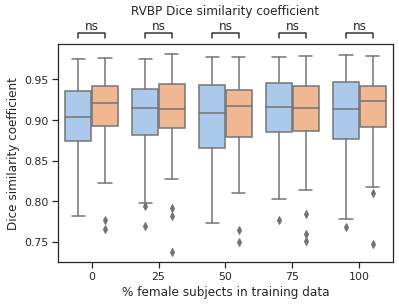}
    \caption{}\label{fig:SexRVBP}
\end{subfigure}
    \hfill
\begin{subfigure}{.3\linewidth}
    \centering
    \includegraphics[scale=0.3]{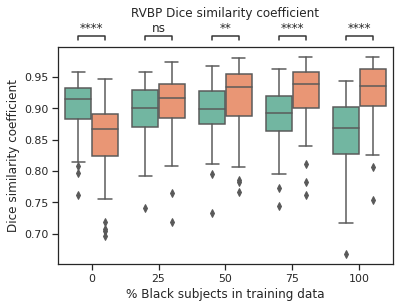}
    \caption{}\label{fig:BlackRVBP}
\end{subfigure} 
    \hfill
\begin{subfigure}{.3\linewidth}
    \centering
    \includegraphics[scale=0.3]{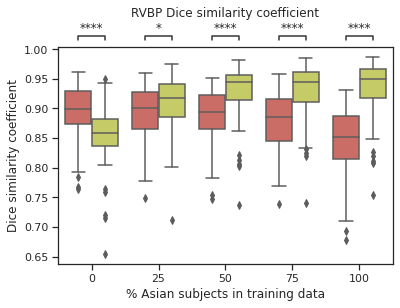}
    \caption{}\label{fig:AsianRVBP}
\end{subfigure}

\caption{Comparison of DSC for the left ventricular blood pool (LVBP) (a-c), left ventricular myocardium (LVM) (d-f), and right ventricular blood pool (RVBP) (g-i) for the three experiments. Statistical significance was found using a Mann-Whitney U test and is denoted by **** \((p \leq 0.0001)\), *** \((0.001 < p \leq 0.0001)\), ** \((0.01 < p \leq 0.001)\), * \((0.01 < p \leq 0.05)\), ns \(( 0.05 \leq p)\).}
\label{fig:regions_boxplots}
\vspace{-3mm}
\end{figure*}

\subsection{Experiment 3: White \vs Asian}
\cref{tab:average DSC} shows that the highest overall median DSC for Experiment 3 was achieved when the dataset was comprised of 75\% Asian subjects and 25\% white subjects. \cref{fig:AsianDSC} shows a similar trend to the results from Experiment 2. However, in this experiment, with a training dataset that is 100\% Asian, larger differences can be observed between the white and Asian subjects than were observed for the white and black subjects. For both race experiments, the largest differences in DSC scores between the groups were found in the LVM (\cref{fig:BlackLVM} and \cref{fig:AsianLVM}). 

Accuracy parity was also achieved here with a split of 75\% white subjects and 25\% Asian patients. Increasing the proportion of Asian subjects from 0\% to 25\% increased the DSC for Asian subjects from 0.912 to 0.955 (p$<$0.0001) and the overall DSC from 0.930 to 0.949 (p$<$0.0001) but did not increase the DSC for the white subjects. Further increasing the proportion of Asian subjects to 50\% significantly increased the DSC for the Asian subjects to 0.972 (0.01$<$ p$<$0.001) and the overall DSC to 0.954.

\begin{figure*}
\centering

\begin{subfigure}{.3\linewidth}
    \centering
    \figuretitle{Experiment 1}
    \includegraphics[scale=0.3]{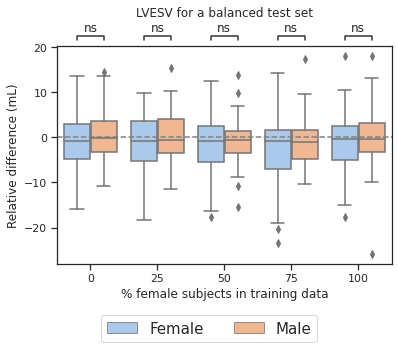}
    \caption{}
    \label{fig:SexLVSV}
\end{subfigure}
    \hfill
\begin{subfigure}{.3\linewidth}
    \centering
    \figuretitle{Experiment 2}
    \includegraphics[scale=0.3]{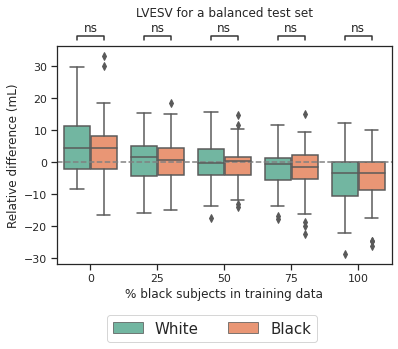}
    \caption{}
    \label{fig:BlackLVSV}
\end{subfigure}
    \hfill
\begin{subfigure}{.3\linewidth}
    \centering
    \figuretitle{Experiment 3}
    \includegraphics[scale=0.3]{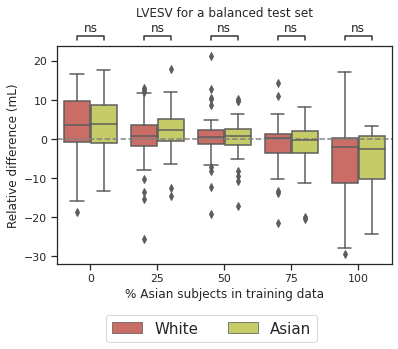}
    \caption{}
    \label{fig:AsianLVSV}
\end{subfigure}

\bigskip
\begin{subfigure}{.3\linewidth}
    \centering
    \includegraphics[scale=0.3]{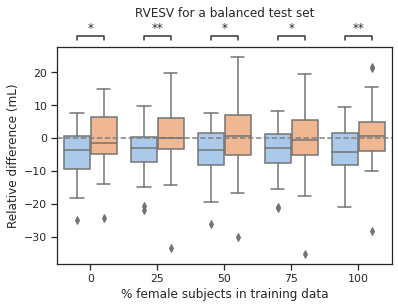}
    \caption{}\label{fig:SexRVSV}
\end{subfigure}
    \hfill
\begin{subfigure}{.3\linewidth}
    \centering
    \includegraphics[scale=0.3]{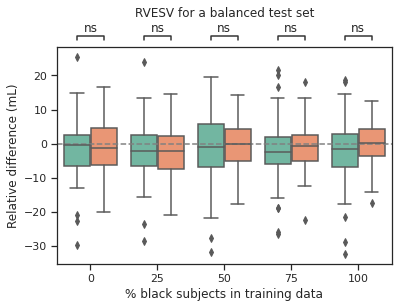}
    \caption{}\label{fig:BlackRVSV}
\end{subfigure} 
    \hfill
\begin{subfigure}{.3\linewidth}
    \centering
    \includegraphics[scale=0.3]{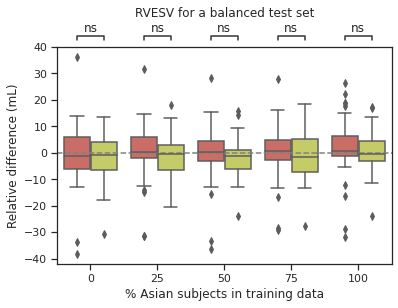}
    \caption{}\label{fig:AsianRVSV}
\end{subfigure}

\bigskip
\begin{subfigure}{.3\linewidth}
    \centering
    \includegraphics[scale=0.3]{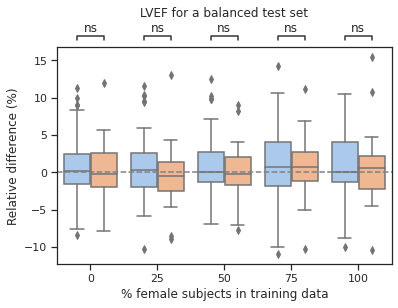}
    \caption{}\label{fig:SexLVEF}
\end{subfigure}
    \hfill
\begin{subfigure}{.3\linewidth}
    \centering
    \includegraphics[scale=0.3]{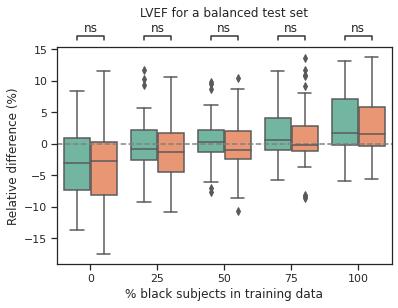}
    \caption{}\label{fig:BlackLVEF}
\end{subfigure} 
    \hfill
\begin{subfigure}{.3\linewidth}
    \centering
    \includegraphics[scale=0.3]{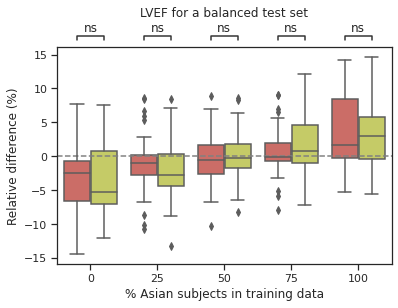}
    \caption{}\label{fig:AsianLVEF}
\end{subfigure}

\bigskip
\begin{subfigure}{.3\linewidth}
    \centering
    \includegraphics[scale=0.3]{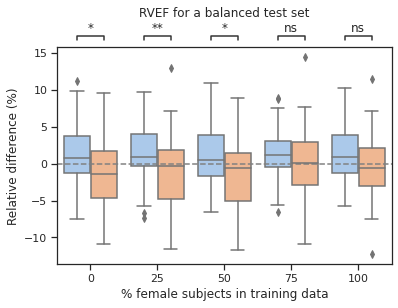}
    \caption{}\label{fig:SexRVEF}
\end{subfigure}
    \hfill
\begin{subfigure}{.3\linewidth}
    \centering
    \includegraphics[scale=0.3]{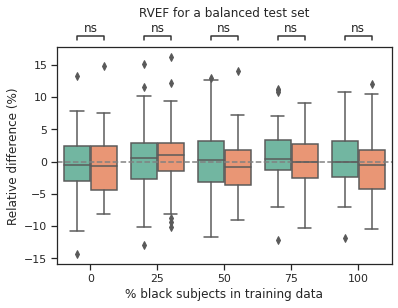}
    \caption{}\label{fig:BlackRVEF}
\end{subfigure} 
    \hfill
\begin{subfigure}{.3\linewidth}
    \centering
    \includegraphics[scale=0.3]{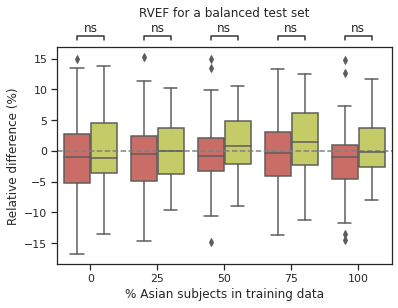}
    \caption{}\label{fig:AsianRVEF}
\end{subfigure}

\caption{Relative difference in LV and RV end-systolic volume (ESV) and ejection fraction (EF) for the three experiments. The differences were calculated by subtracting ground truth values from predicted values e.g. $LVESV_{difference} = LVESV_{predicted} - LVESV_{ground\text{ } truth}$. Statistical significance was found using a Mann-Whitney U test and is denoted by ** \((0.01 < p \leq 0.001)\), * \((0.01 < p \leq 0.05)\), ns \(( 0.05 \leq p)\).}
\label{fig:clinical_measures}
\end{figure*}

The results for the clinical measures (\cref{fig:clinical_measures}) for this experiment show similar findings to those for Experiment 2.

\begin{table}
  \caption{Intersectional analysis of median DSC scores broken down by the protected attributes used in the experiments. The train percentages signify the proportions of each protected group used in training, i.e. female/male for Experiment 1, black/white for Experiment 2 and Asian/white for Experiment 3. Statistical significance was found using a Mann-Whitney U test and is denoted by *** \((0.001 < p \leq 0.0001)\), ** \((0.01 < p \leq 0.001)\), * \((0.01 < p \leq 0.05)\).}
  \centering
  \resizebox{\textwidth}{!}{
  \begin{tabular}{c|cc|cc|cc|cc|cc|cc}
    \toprule
     &  \multicolumn{4}{c|}{Experiment 1}  & \multicolumn{4}{c|}{Experiment 2} &  \multicolumn{4}{c}{Experiment 3} \\
     Train &  \multicolumn{2}{c}{Female}  & \multicolumn{2}{c|}{Male} &  \multicolumn{2}{c}{Black} & \multicolumn{2}{c|}{White} & \multicolumn{2}{c}{Asian} & \multicolumn{2}{c}{White}  \\
    Percentage & White & Black & White & Black & Male & Female & Male & Female & Male & Female & Male & Female\\
     \midrule
    0\%/100\% & 0.950 & 0.963 ** & 0.948 & 0.963 ** & 0.922 & 0.925 & 0.948 & 0.954 & 0.919 & 0.909 & 0.945 & 0.944 \\
    25\%/75\% & 0.952 & 0.962 * & 0.955 & 0.969 * & 0.951 & 0.959 & 0.956 & 0.952 & 0.952 & 0.960 & 0.944 & 0.944 \\
    50\%/50\% & 0.951 & 0.968 *** & 0.957 & 0.962 & 0.961 & 0.967 & 0.945 & 0.953 & 0.970 & 0.973 & 0.945 & 0.944 \\
    75\%/25\% & 0.952 & 0.970 ** & 0.951 & 0.965  *  & 0.964 & 0.971 & 0.952 & 0.949 & 0.970 & 0.975 & 0.934 & 0.937 \\
    100\%/0\% & 0.954 & 0.969 ** & 0.950 & 0.971  ** & 0.971 & 0.969 & 0.936 & 0.913 & 0.965 & 0.975 & 0.901 & 0.923\\
    \bottomrule
  \end{tabular}}
  \label{tab:intersectional_results}
\end{table}

\section{Discussion}
\label{sec:discussion}
To the best of our knowledge, this work has presented the first systematic study of how imbalances in training datasets affect AI-based CMR segmentation performance. Our results show that a significant bias towards the majority group can be seen when the datasets are racially imbalanced. We did not find any significant sex bias.
\newline \indent  As discussed in \cref{sec:results}, the median DSC for both the black and Asian subjects increased significantly when their representation in the dataset was increased from 0\% to 25\%. It could be argued that a model trained with this dataset is more fair than a model trained with 100\% white subjects as the DSC for the minority subjects increased while the DSC for the white subjects remained approximately the same. Increasing the proportion of the minority races to 50\% further increased the overall DSC scores, suggesting that a more diverse training set will produce a model with a better segmentation performance. In clinical applications, cardiac segmentations are used to obtain important measures of cardiac function such as left- and right-ventricular ES volume and ED volume (see \cref{fig:clinical_measures}), which are used for diagnosis, prognosis and treatment planning for patients \cite{Lekadir2020DeepLearning}. Given that the prevalence of cardiovascular diseases is known to be higher in minority races, producing models which are fairer overall is essential to improve patient management and reduce healthcare inequalities.
\newline \indent However, we note that if the tasks of segmenting CMR images of different races had equal difficulty, we would expect the dataset with a 50\%/50\% split to have equal DSC scores for both races. In Experiment 2, with a balanced dataset, the black subjects achieved a DSC score which was higher than that for white subjects. The same pattern can be seen for Experiment 3 considering Asian subjects.
This suggests that there may be more `within-group' variation in the anatomies of the white subjects than in the black or Asian subjects, making segmenting white subjects' hearts an inherently harder problem for the AI model to solve. Applying bias mitigation techniques such as those introduced in \cite{Puyol-Anton2021FairnessSegmentation} may produce a model which does not generalise well to anatomical differences between protected groups which may carry important diagnostic information. Given that a subject's self-identified race is (likely) available at inference time, using a model specific to the subject's protected group may prove to be more fair than using a single model for all subjects.
\newline \indent Analysis of the clinical measures (\cref{fig:clinical_measures}) revealed some interesting and perhaps surprising findings. There was no clear trend in relative differences in clinical measures, even when there was such a trend in DSC in Experiments 2 and 3. Furthermore, the lowest relative differences for a protected group did not always occur when that group was in the majority in the training set. For example, consider \cref{fig:BlackLVEF} which shows that the lowest relative difference in LVEF for white subjects does not occur when the training set is 100\% white, but rather the lowest errors for both black and white subjects occur when the training set is balanced. These findings require further investigation, but add further weight to our argument that considering training set (im)balance is crucial for training fair and robust AI-based segmentation models.
\newline \indent Nevertheless, in terms of spatial overlap measures, our results have shown clear racial biases in segmentation performance. However, it is possible that race is not the underlying factor that explains the differences in performance observed, and the differences could instead be caused by a confounding factor. In our previous work \cite{10.3389/fcvm.2022.859310} we investigated a range of potential confounders and discovered none that could explain the bias. We remain open-minded about the existence of such confounders, and future work will investigate the effect of imbalances of other protected attributes such as socioeconomic status and age.
We also emphasise that racial bias in CMR segmentation is a cause for concern regardless of its underlying explanation.
\newline \indent Overall, including more minorities increased overall DSC scores and scores for individual minority groups. Therefore, we advocate for the inclusion of more minorities in public medical datasets and the transparent reporting of performance of AI models separately for all protected groups.
Although varying the proportion of females in the dataset did not result in significant differences in segmentation performance, it is essential that females have equal representation in publicly available medical datasets as previous work discussed in \cref{sec:intro} has found that AI-based diagnosis methods frequently underperform for females.
This work has highlighted the need for the fair representation of minority groups in medical imaging datasets.
To the best of our knowledge, this work is the first to systematically investigate the effect of dataset imbalances on segmentation accuracy.

\section*{Acknowledgements}
This work was supported by the Engineering \& Physical Sciences Research Council  Doctoral Training Partnership (EPSRC DTP) grant  EP/T517963/1. This research has been conducted using the UK Biobank Resource under Application Number 17806.

\printbibliography

\end{document}